\documentclass[aps,prx,twocolumn,floatfix,superscriptaddress,amsmath,amssymb]{revtex4-1}

\usepackage{graphicx}
\usepackage{dcolumn}%
\usepackage{bm}
\usepackage[pdftex]{color}
\usepackage[dvipsnames]{xcolor}
\usepackage[colorlinks,bookmarks=false,citecolor=RoyalBlue,linkcolor=BrickRed,urlcolor=OliveGreen]{hyperref}
\usepackage[mathlines]{lineno}
\usepackage{braket}
\usepackage{bbm}
\usepackage{verbatim}
\usepackage{ulem}
\usepackage{soul}

\newcommand{\tperi}{\text{peri}}
\newcommand{\tarea}{\text{area}}

\def\ie{\begin{equation}\begin{aligned}}
\def\fe{\end{aligned}\end{equation}}

\begin{document}

\newcommand{\thetitle}{Finite Correlation Length Scaling of Disorder Parameter at Quantum Criticality}
\title{\thetitle}

\newcommand{\ugent}[0]{Department of Physics and Astronomy, University of Ghent, 9000 Ghent, Belgium}

\author{Wen-Tao Xu}
\affiliation{Technical University of Munich, TUM School of Natural Sciences, Physics Department, 85748 Garching, Germany}
\affiliation{Munich Center for Quantum Science and Technology (MCQST), Schellingstr. 4, 80799 M{\"u}nchen, Germany}

\author{Rui-Zhen Huang}
\email{huangrzh@icloud.com}
\affiliation{Graduate School of China Academy of Engineering Physics, Beijing 100193, China}
\affiliation{\ugent}

\date{\today}

\begin{abstract}
The disorder parameter, defined as the expectation value of the symmetry transformation acting on a subsystem, can be used to characterize symmetric phases as an analogy to detecting spontaneous symmetry breaking (SSB) phases using local order parameters. In a dual picture, disorder parameters actually detect SSB of higher-form symmetries. In this work, we show that the non-local disorder parameters can be conveniently and efficiently evaluated using infinite projected entangled pair states (iPEPS). Moreover, we propose a finite correlation length scaling theory of the disorder parameter within the quantum critical region and validate the scaling theory with variationally optimized iPEPS. We find from the finite \textcolor{black}{correlation length} scaling that the disorder parameter satisfies perimeter law at a critical point, i.e., it decays exponentially with the boundary size of the subsystem, indicating spontaneous higher-form symmetry breaking at the critical point of the dual model.   
\end{abstract}

\maketitle
\textbf{Introduction.} Conventional phases and phase transitions are described by the \textcolor{black}{Landau-Ginzburg-Wilson} paradigm~\cite{landau2013statistical,wilson1974renormalization}, where the local order parameters play an essential role, because a non-zero local order parameter implies a spontaneous symmetry-breaking (SSB) phase with a long-range order. Actually, phase and phase transitions can also be characterized by a non-local disorder parameter~\cite{Kadanoff_1971}. For a quantum system with a global on-site symmetry, the disorder parameter is the expectation value of the disorder operator, which is the on-site symmetry transformation applying on an enough large subsystem~\cite{Fradkin2017}. For instance, in one (two) spatial dimensions, the disorder parameter is an expectation value of a string-like (membrane-like) operator. Moreover, by decorating the boundary of disorder operators, we can detect non-trivial symmetry-protected (enriched) topological phases~\cite{SOP_2008,Pollmann_SOP_2012,Jutho_SOP_2012,Pollmann_MOP_2014}. So, disorder parameters play an important role to characterize various symmetric phases.  

On the other hand, disorder parameters are deeply connected to higher-form symmetries~\cite{Zohar_2004,NUSSINOV_2009,Hastings_and_Wen_2005,Higher_sym_gauge_kuasptin_2015,High_form_Kapistin_2015,High_form_wen_2019,Wen_emergent_high_form_2023,Self_dual_critically_Ising,McGreevy_2023}. The $p$-form symmetry transformation of a quantum system in $d$ spatial dimensions applies on a $(d-p)$-dimensional sub-manifold of the system. For instance, the usual global on-site symmetry is a $0$-form symmetry in any dimensions, and the Wilson loop symmetry of the $2d$ toric code model~\cite{kitaev_2002} is a 1-form symmetry. Via a generalized duality transformation~\cite{Wegner_duality_1971,Trebst_2007,Levin_Gu_2012}, which is closely related to ``gauging"~\footnote{The degrees of freedom of the original system can be viewed as matter degrees of freedom. After gauging the symmetry $G$, we obtain a gauge theory with both gauge and matter degrees of freedom which are constrained by the Gauss law, and an isometry transformation can be applied to remove the matter degrees of freedom and the Gauss law constraint~\cite{Tupitsyn_2010,xu_2024_ent_gauge}. The combination of gauging and the isometry transformation is equivalent to the generalized duality transformation.}, a quantum system with a global symmetry $G$ is mapped to a dual quantum lattice gauge system with a $(d-1)$-form symmetry~\cite{Ham_formulation_1975}. The $G$ symmetric (symmetry-breaking) phase of the original model is mapped to a $(d-1)$-form symmetry-breaking (symmetric) phase of the dual lattice gauge model, and the disorder parameter of $G$ is mapped to the order parameter of the $(d-1)$-form symmetry. From the duality to lattice gauge models, one can find that in the $G$ SSB phase of the original model, the disorder parameter decays exponentially with the size of subsystem, the so-called area law. If the disorder parameter decays exponentially with the boundary size of the subsystem, the so-called perimeter law, the original model is in the $G$ symmetric phase~\cite{Wegner_duality_1971,Kogut_1979}. More detailed and precise descriptions of the relation between higher-form symmetries and the disorder parameters can be found in the supplementary material (SM)~\cite{appendix}. 
Since spontaneous higher-form symmetry breaking gives rise to topological orders~\cite{High_form_Kapistin_2015,High_form_wen_2019}, the study of disorder parameters is not only a different viewpoint to understand conventional quantum phases and phase transitions, but also provides information for topological phases and topological phase transitions.

In one-dimensional quantum systems, local order parameters and disorder parameters are dual with each other under the Kramers-\textcolor{black}{Wannier} duality transformation\textcolor{black}{~\cite{KW_1941,Wegner_duality_1971}}, hence at self-dual quantum critical points, both local order parameters and disorder parameters are zero, indicating that both original models and dual models are symmetric. This is also true for systems with non-invertible symmetries in terms of matrix product operators (MPOs)~\cite{Xu_2021,Xu_2022}. However, in two-dimensional quantum systems, the local order parameter and the disorder parameter are no longer dual with each other, and previous study based on the quantum Monte Carlo (QMC) method suggests that at a two-dimensional quantum critical point~\cite{Meng_disorder_para_2021}, the disorder parameter satisfies the perimeter law, implying that the 1-form symmetry is spontaneously broken in the dual gauge model.

\begin{figure}[tbp]
\centering
\includegraphics{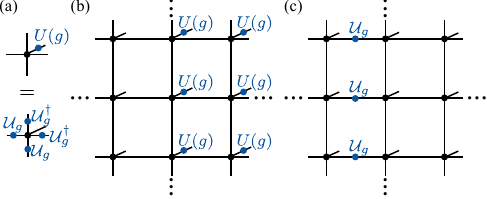}
\caption{(a) In the (0-form) $G$ symmetric topologically trivial phase, the symmetry action on physical indices of the iPEPS tensor can be transformed to the virtual indices. (b) We consider a non-local disorder operator which is the symmetry transformation acting on the right half of the iPEPS. (c) Using the relation in (a), the disorder operator on the physical level of the iPEPS is transformed into a string operator at the virtual level of the iPEPS.}
\label{Fig:PEPS}
\end{figure}

However, in the QMC method, it is not convenient to calculate the disorder parameter since one needs to extrapolate with the subsystem size,  and it is inefficient to calculate the disorder parameter in the $G$ SSB phase~\cite{Meng_disorder_para_2021}, \textcolor{black}{especially in systems that are affected by the sign problem}. An alternative approach is tensor network states (TNS)~\cite{maeshima:2001,verstraete2004}, i.e., infinite matrix product state (iMPS) in one dimension and infinite projected entangled pair states (iPEPS) in two dimensions, which provide an alternative powerful framework to study quantum phases and phase transitions.  In this work, we first show that by using iPEPS, one can directly calculate the coefficients of the area law and the perimeter law of the disorder parameter without any extrapolation. Moreover, distinct from the intuition that TNS can only characterize gapped phases, it was found that TNS can be used to extract universal data at quantum criticality via finite correlation length scaling~\cite{Pollmann_2009,Corboz_Finite_IPEPS_2018,Lauchli_Finite_IPEPS_2018,Bram_2019,Bram_2022}~\textcolor{black}{, also known as finite entanglement scaling in the literature}. In one dimension, it was suggested the the entanglement scaling is related to a perturbative conformal field theory~\cite{hrz_2024}. Nevertheless, previous studies based on the \textcolor{black}{finite correlation length} scaling focused on local physical observables. In two-dimensional systems, disorder parameters are non-local observables. We generalize the \textcolor{black}{finite correlation length} scaling to disorder parameters and propose a scaling theory to the area law and perimeter law coefficients of disorder parameters. At quantum critical points, using the variational iPEPS simulations, we validate the scaling theory and examine the fate of the area law coefficients, which dominate the behavior of the disorder parameters.  

\textbf{Disorder parameter from iPEPS.} It is believed that TNS can efficiently approximate ground states of gapped (non-chiral) quantum many-body systems~{\textcolor{black}{\footnote{It is believed that there is a no-go theorem which claims that iPEPS can not efficiently represent gapped chiral quantum states~\cite{Chiral_PEPS_no_go,Chiral_PEPS_tu}.}}. An iPEPS defined on the square lattice is a network consisting of rank-5 tensors arranged repeatedly, as shown in Fig.~\ref{Fig:PEPS}. We can show that the disorder parameter satisfies the perimeter law in symmetric gapped phases and area law in gapped SSB phases using iPEPS, and the area and perimeter law coefficients can be directly extracted from iPEPS. 

Consider a system with a global on-site symmetry $G$: $\prod_iU_i(g)$, where the unitary representation $U_i(g)$ with $g\in G$ acts on the site $i$. For simplicity, we consider the bi-partition of an infinite lattice into the left part $L$ and the right part $R$, and the disorder operator is defined as $\prod_{i\in R}U_i(g)$. Using an iPEPS $\ket{\Psi}$, the disorder parameter $\tilde{m}$ can be expressed as:
\begin{equation}\label{eq:disorder_PEPS}
     \tilde{m}_g = \frac{\bra{\Psi}\prod_{i \in R} U_i(g) \ket{\Psi}}{\langle\Psi|\Psi\rangle}= \lim_{N_x \rightarrow \infty} \frac{ \mathrm{Tr} \left( T^{N_x/2}T_g^{N_x/2} \right)}{\mathrm{Tr} \left( T^{N_x} \right) },
\end{equation}
where $T$ and $T_g$ are iPEPS transfer matrices defined in Fig.~\ref{Fig:transfer_operator}(b). For a $G$ symmetric iPEPS (without non-trivial symmetry protected/enriched topological order), $U(g)$ applying on the physical indices of the rank-5 iPEPS tensor can be transformed to the virtual indices~\cite{RMP_MPS_PEPS_2021}, as shown in Fig.~\ref{Fig:PEPS}(a), and the disorder operator applying on $R$ can be transformed from the physical level to virtual level of the iPEPS~\cite{Iqbal_2021}, as shown in Figs.~\ref{Fig:PEPS}(b) and (c), thus the disorder parameter can be simplified to:
\textcolor{black}{\begin{equation}\label{eq:dis_op_sym_1}
\tilde{m}_g=\lim_{N_y\rightarrow\infty}\frac{\langle V \vert \mathbbm{1}^{\otimes N_y}\otimes \mathcal{U}_g^{\otimes N_y} \vert V\rangle}{\langle V |V\rangle},
\end{equation}}
where \textcolor{black}{$\mathbbm{1}$ is the identity operator on the virtual degrees of freedom of the bra iPEPS $\bra{\Psi}$, and $\mathcal{U}_g$ applies on the virtual degrees of freedom of the ket iPEPS $\ket{\Psi}$}, $\ket{V}$ is the \textcolor{black}{(unnormalized)} leading eigenvector of $T$ (Fig.~\ref{Fig:transfer_operator}(b)), and $\mathcal{U}_g$ is defined in Fig.~\ref{Fig:PEPS}(a). Here we assume all the transfer matrices are Hermitian for simplicity, and the generalization to the non-Hermitian case is straightforward. When the iPEPS is gapped (with a finite correlation length), $\ket{V}$ can be efficiently represented by an iMPS, and the calculation of $\tilde{m}_g$ can be further expressed in terms of the 0-dimensional transfer matrices $\tilde{\mathcal{T}}_g$ and $\mathcal{T}$ shown in Fig.~\ref{Fig:transfer_operator}(c), such that the disorder operator $\tilde{m}$ can be finally expressed as:
\begin{equation}\label{eq:dis_op_sym_2}
    \tilde{m}_g =\lim_{N_y\rightarrow\infty} \frac{\tilde{\lambda}_g^{N_y}}{\lambda^{N_y}}=\lim_{N_y\rightarrow\infty}\exp(-\alpha_{\tperi} N_y),
\end{equation}
where $\tilde{\lambda}_g$ and $\lambda$ are dominant eigenvalues of $\tilde{\mathcal{T}}_g$ and $\mathcal{T}$ respectively, from which the perimeter law coefficient can be directly extracted: $\alpha_{\tperi}=\ln(\lambda/\tilde{\lambda}_g)$~\footnote{Furthermore, using dominant eigenvectors of $\tilde{\mathcal{T}}_g$ and $\mathcal{T}$, one can calculate the constant correction $\gamma$ to the perimeter law in topological phases~\cite{Xu_MOP_2023}: $e^{-\alpha_{\tperi}N_y+\gamma}$, which is the so-called topological disorder parameter~\cite{Topo_disorder_para_2022}}. 
So we conclude that the disorder parameter of the $G$ symmetric gapped iPEPS always satisfies the perimeter law, indicating that in the dual lattice gauge model, the 1-form symmetry is broken spontaneously. In addition, this indicates that the disorder parameter of the iPEPS constructed from a two-dimensional classical partition function~\cite{PEPS_from_classical_2006} always satisfies the perimeter law even if the classical partition function is in a SSB phase~\cite{Gaining_insight_2023}.  

In the $G$ SSB phase, iPEPS tensors no longer satisfy the relation shown in Fig.~\ref{Fig:PEPS}(a). The disorder operator can not be transformed to the virtual level of the iPEPS. However, the disorder parameter can still be calculated using transfer matrices:
\begin{align}\label{Eq:dis_op_from_transfer_matrix}
    \tilde{m}_g &= \lim_{N_x,N_y \rightarrow \infty} \frac{\langle V|V_{g}\rangle}{\sqrt{\langle V|V\rangle \langle V_{g}|V_{g}\rangle}} \left( \frac{t_g}{t} \right) ^{N_x N_y/2}\notag\\
    &=\lim_{N_x,N_y \rightarrow \infty} \left( \frac{\lambda_{g,\mathrm{ovlp}}}{\sqrt{\lambda\lambda_g}} \right) ^{N_y}\left( \frac{t_g}{t} \right) ^{N_x N_y/2},
\end{align}
where $t$ \textcolor{black}{($V$)} and $t_g$ \textcolor{black}{($V_g$)} are the leading eigenvalue  \textcolor{black}{densities (unnormalized leading eigenvectors)} of the transfer matrix $T$ and $T_g$ shown in Fig.~\ref{Fig:transfer_operator}(b), respectively, and $\lambda$, $\lambda_g$ and $\lambda_{g,\mathrm{ovlp}}$ are dominant eigenvalue of $0d$ transfer matrix $\mathcal{T}$, $\mathcal{T}_g$ and $\mathcal{T}_{g,\mathrm{ovlp}}$ defined in Fig.~\ref{Fig:transfer_operator}(c) respectively. 
From Eq.~\eqref{Eq:dis_op_from_transfer_matrix}, we conclude that the disorder parameter of iPEPS in the gapped $G$ SSB phase has the form:
\begin{equation} \label{eq_disorder_ssb}
    \tilde{m}_g \sim e^{-\alpha_{\tperi} \vert \partial R \vert - \alpha_{\tarea} \vert R \vert },
\end{equation}
where $|R|=N_xN_y/2$ and $|\partial R|=N_y$ and $N_x,N_y\rightarrow+\infty$, $\alpha_{\tperi}=\ln(\sqrt{\lambda\lambda_g}/\lambda_{g,\mathrm{ovlp}})$ and \textcolor{black}{$\alpha_{\tarea}=\ln(t/t_g)>0$}~\footnote{When the iPEPS is $G$ symmetric, $t=t_g$ $(\alpha_{\text{area}}=0)$, $\lambda=\lambda_g$ and $\tilde{\lambda}_g=\lambda_{g,\text{ovlp}}$,
and it reduces back to the perimeter law case we discussed in Eqs.~\eqref{eq:dis_op_sym_1} and~\eqref{eq:dis_op_sym_2}}.  Compared to QMC method, we can directly extract the perimeter law coefficient $\alpha_{\tperi}$ and the area law coefficient $\alpha_{\tarea}$ from iPEPS without extrapolation, and iPEPS do not have the inefficiency of calculating the disorder parameter in SSB phases using QMC methods.

\begin{figure}[tbp]
\centering
\includegraphics{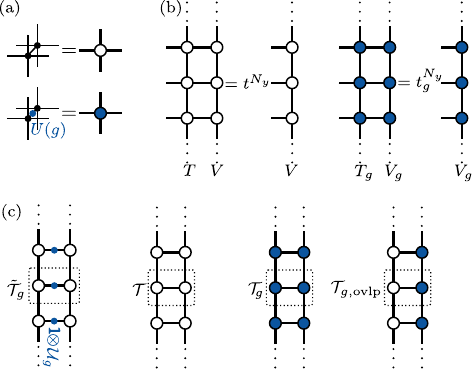}
\caption{(a) The iPEPS double tensor is defined by contracting the physical indices of two iPEPS tensors, and it is reshaped by combining the corresponding virtual indices in the bra and ket layers. We also consider another double tensor where an $U(g)$ operator is sandwiched between iPEPS tensors in the bra and ket layers. (b) From the double tensors we can construct the transfer matrices and approximate their fixed points using iMPS, $t$ and $t_g$ are density of the largest eigenvalues. (c) From the iMPS, various $0d$ transfer matrices can be defined.}
\label{Fig:transfer_operator}
\end{figure}

\begin{figure*}[tbp]
\centering
\includegraphics[scale=0.5]{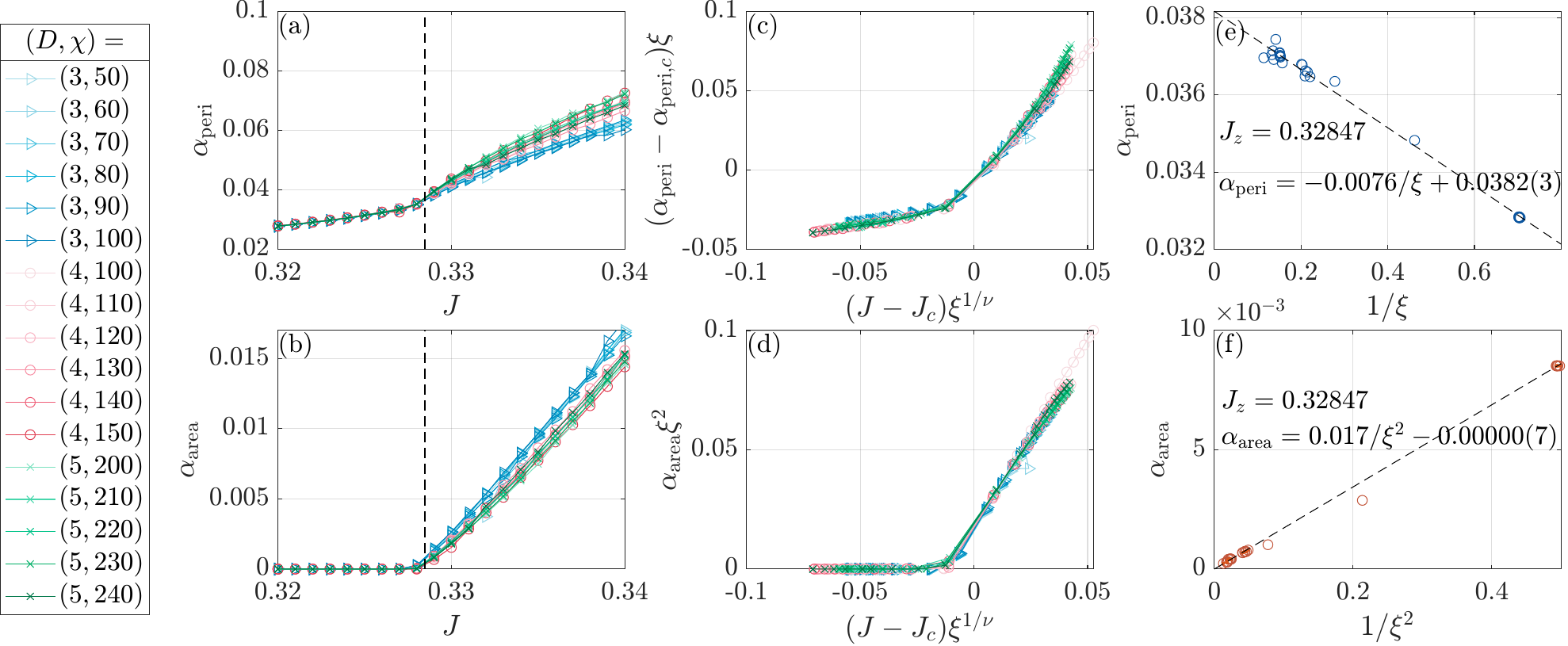}
\caption{Result of the $(2+1)d$ quantum Ising model. (a) and (b) The perimeter law coefficient $\alpha_{\tperi}$ and area law coefficient $\alpha_{\tarea}$ obtained from iPEPS with various bond dimensions $(D,\chi)$, separately. (c) and (d) Extrapolating $\alpha_{\tperi}$ and $\alpha_{\tarea}$ at the critical point $J_c\approx 0.32487$ to the limit $\xi\rightarrow\infty$ separately, where the data are obtained from iPEPS with bond dimensions $D=2,3,4,5$ and $\chi$ ranging from $5$ to $250$, and we only show the uncertainty from the extrapolation. (e) Data collapse of $\alpha_{\tperi}$ in (a), where we use $\alpha_{\tperi,c}$ obtained from (c) and $\nu=1/(d+z-\Delta_y)= 0.629970(4)$~\cite{Conformal_bootstrap_2016}. (f)  Data collapse of $\alpha_{\tarea}$ in (b).}
\label{Fig:results_in_main_text}
\end{figure*}

\textbf{\textcolor{black}{Finite correlation length} scaling of disorder parameter at criticality.}
In the quantum critical region, the emergent scale invariance is captured by the finite length scaling analysis. In the iMPS/iPEPS framework, the most convenient way is to apply the \textcolor{black}{finite correlation length} scaling~\cite{Pollmann_2009,Corboz_Finite_IPEPS_2018,Lauchli_Finite_IPEPS_2018,Bram_2019,Bram_2022}. At the critical points described by CFTs, the correlation length diverges and the ground state can not be efficiently expressed as iMPS or iPEPS with a finite bond dimension $D$, which induces a finite correlation length $\xi(D)$. However, by increasing the bond dimension $D$, physical observables exhibit universal scaling behavior with $\xi(D)$. An interesting observation is that at the critical point, the iMPS or iPEPS with a finite $D$ tends to spontaneously break the global symmetry, because the finite bond dimension truncation induces a symmetry allowed relevant perturbation~\cite{hrz_2024}. In contrast, the ground state of a finite-size system usually preserves the global symmetry at the critical point.

We first generalize the scaling theory of local physical observables to disorder parameters. Notice that there is only one length scale $\xi(D,\chi)$ in an iPEPS~\cite{Bram_2022}, where $\chi$ is the bond dimension of the iPEPS environment tensors. By considering the general form of the disorder parameter in gapped phases in Eq.~\eqref{eq_disorder_ssb} and replacing the area $|\partial R|$ ($|R|$) with $\xi$ ($\xi^{2}$), one can write down the expression for disorder parameters near critical points for a given length scale $\xi$:
\begin{equation}\label{eq_scaling_wrong}
  \tilde{m}_g \sim e^ {- \xi\,\alpha_{\tperi} - \xi^2 \, \alpha_{\tarea}}.  
\end{equation}
The expression of the disorder parameter $\tilde{m}_g$ indicates that it is a dimensionless quantity, and $\alpha_\text{peri}$ and $\alpha_\text{area}$ have the scaling dimensions $\Delta(\alpha_{\tperi}) = 1$ and $\Delta(\alpha_{\tarea}) = 2$ respectively. Accordingly, we can write down a scaling theory:
\ie \label{eq_scaling}
&\alpha_{\tperi}(y, \xi) \sim \alpha_{\tperi}(y) + \xi^{-1} f_0(y\xi^{d+z-\Delta_y}), \\  
&\alpha_{\tarea}(y, \xi) \sim \alpha_{\tarea}(y) + \xi^{-2} f_1(y\xi^{d+z-\Delta_y}),
\fe
where $y$ is a relevant parameter, $z$ is the dynamical critical exponent, $\Delta_y$ is the scaling dimension of the relevant field coupled to $y$, and $f_0$ and $f_1$ are dimensionless scaling functions. 

At a quantum critical point, i.e., $y=0$, the global symmetry must be preserved and the local order parameter becomes zero. However, as we mentioned, with a finite entanglement cutoff, the variationally optimized iPEPS at the critical point is actually in the SSB phase. As a result, the disorder parameter satisfies area law, indicating that in the dual lattice gauge model, the 1-form symmetry does not break. It would be interesting to study whether the disorder parameter becomes the perimeter law in the infinite $\xi$ limit at which the broken global symmetry $G$ in iPEPS is recovered, i.e., consider whether $\lim_{\xi\rightarrow\infty}\alpha_{\tarea}(y=0,\xi)$ becomes zero. Notice that in the QMC method, the symmetry $G$ does not break at the critical point, which is essentially different from iPEPS, so studying the behavior of the disorder parameter at the critical point using iPEPS is an important complementary to the previous QMC study~\cite{Meng_disorder_para_2021}.

\textbf{\textcolor{black}{Finite correlation length} scaling and vanishing of area law coefficient at the Ising criticality.}   
We first consider the disorder parameter of the quantum Ising model on the square lattice:
\begin{equation}\label{Eq:Ising_Ham}
    H_{\text{Ising}}=-J \sum_{\langle i,j\rangle}\sigma^z_i \sigma^z_{j} - h \sum_i \sigma^x_i,
\end{equation}
where $\sigma^z$ and $\sigma^x$ are Pauli matrices and $h=1$ is assumed in the following. The model has a global $\mathbb{Z}_2$ spin flip symmetry generated by $\prod_i\sigma^x_i$. There is a continuous quantum critical point described by the $(2+1)d$ Ising CFT at $J_c = 0.328474(3)$~\cite{Youjin_2002}, which is between the gapped $\mathbb{Z}_2$ symmetric phase ($J<J_c$) and the gapped $\mathbb{Z}_2$ SSB ($J>J_c$) phase. We use the gradient-based variational iPEPS to optimize the ground states~\cite{Laurens_gradient,iPEPS_corboz_2016}. The environment for the iPEPS is calculated using the corner transfer matrix renormalization group (CTMRG)~\cite{CTMRG_1,CTMRG_2} with a bond dimension $\chi$, and the energy density gradient is obtained through automatic differentiation~\cite{Liao_AD_2019}.

Figures.~\ref{Fig:results_in_main_text}(a) and (b) show the perimeter law coefficient $\alpha_{\tperi}$ and the area law coefficient $\alpha_{\tarea}$ of the disorder parameter $\langle \prod_{i\in R} X_i\rangle$ obtained from iPEPS with various bond dimensions $(D,\chi)$ \textcolor{black}{using Eqs.~\eqref{Eq:dis_op_from_transfer_matrix} and ~\eqref{eq_disorder_ssb}}. At the quantum critical point we perform finite \textcolor{black}{correlation length} scaling analysis of $\alpha_{\tperi}$ and $\alpha_{\tarea}$, which show correct scaling dimensions $\Delta(\alpha_{\tperi})=1$ and $\Delta(\alpha_{\tarea})=2$, see \textcolor{black}{Figs.~\ref{Fig:results_in_main_text}(e) and (f)}. We also find the converged value of \textcolor{black}{$\alpha_{\tperi,c}=0.0382$} when the correlation length $\xi\rightarrow +\infty$ in \textcolor{black}{Fig.~\ref{Fig:results_in_main_text}(e)}, which is close to the result $\alpha_{\tperi}=0.0394$ obtained from the QMC simulation~\cite{Meng_disorder_para_2021}. \textcolor{black}{We also show in SM~\cite{appendix} that $\alpha_{\text{peri},c}$ from the two-dimensional quantum Ising model and the three-dimensional classical Ising model are different, indicate $\alpha_{\text{peri},c}$ is a non-universal number.} The area law coefficient $\alpha_{\tarea}$ is more important since it dominates the behavior of the disorder parameter. As we already analyzed above, the global symmetry is indeed broken for finite $D$ iPEPS, as shown by the local order parameter in SM~\cite{appendix}, giving rise to a small but finite $\alpha_{\tarea}$. A careful extrapolation shows that in the limit $\xi\rightarrow +\infty$, $\alpha_{\tarea}$ disappears at the critical point, see \textcolor{black}{Fig.~\ref{Fig:results_in_main_text}(f)}. This demonstrates that the disorder parameter satisfies perimeter law at the Ising critical point, and the 1-form Wilson loop symmetry in the dual $\mathbb{Z}_2$ lattice gauge model --- the toric code model in a longitudinal or transverse field~\cite{Trebst_2007}, breaks spontaneously. Near the critical point, we can collapse the coefficients obtained from different bond dimensions in Figs.~\ref{Fig:results_in_main_text}(a) and (b) on a single curve using Eq.~\eqref{eq_scaling}, as shown in \textcolor{black}{Figs.~\ref{Fig:results_in_main_text}(c) and (d)}, showing that the scaling theory we proposed correctly capture the scaling behavior of $\alpha_{\tperi}$ and $\alpha_{\tarea}$. 

In SM~\cite{appendix}, we also perform the same calculation of the $3d$ classical Ising model, confirming the vanishing of $\alpha_\text{area}$ in the large-length-scale limit and the correct scaling behavior of both $\alpha_\text{area}$ and $\alpha_\text{peri}$. In addition, the Ising criticality we considered here has a dynamical critical exponent $z=1$\textcolor{black}{. We} also show that the scaling of the disorder parameter at a quantum critical point with $z>1$ in SM~\cite{appendix}, at which we find the disorder parameter also satisfies the perimeter law.

\textbf{Disorder parameters at first-order phase transitions.}
To study disorder parameters at first-order phase transitions, we consider the ferromagnetic ($J>0$) $q$-state Potts models on the square lattice:
\begin{equation}\label{Eq:Potts_Ham}
    H_{\text{Potts}}=-J\sum_{\langle i,j\rangle}\sum_{n=1}^{q-1}Z^{n}_i Z^{q-n}_{j}-h\sum_i\sum_{n=1}^{q} X^{n}_i,
\end{equation}
where the clock operator $Z$ and shift operator $X$ defined on the $q$-dimensional space $\{\vert 1\rangle,\vert2\rangle,\dots,\vert q \rangle\}$ are defined as $Z \vert n \rangle = e^{2\pi i n /q} \vert n \rangle$ and $X \, \vert (n+1) \, \mathrm{mod} \ q \rangle = \vert n \rangle$. The $q=2$ model is nothing but the quantum Ising model studied above. It has the global permutation symmetry $S_q$, generated by $\prod_i X_i$ and $\prod_i Y_i$, where $Y\ket{n}=\ket{(-n)\mod q}$ and $Y^2=1$. We consider $h=1$ in the following. By increasing $J$, the model with $q>2$ undergoes a first-order phase transition to an SSB phase where $S_q$ symmetry breaks completely. Using variational iPEPS, we confirmed the first-order transition in the SM~\cite{appendix}. The disorder parameter we consider is defined as $\tilde{m}_Y=\langle \prod_{i\in R}Y_i\rangle$. The area-law coefficients extracted from $\tilde{m}_Y$ of the 4-state and 5-state quantum Potts model have finite jumps across the first-order transition points, as shown in Figs.~\ref{Fig:results_Potts_main_text} (a) and (b), suggesting the coexistence of the area law and the perimeter law at the first-order transitions. This indicates that in the dual lattice gauge models, the 1-form symmetric states and the 1-form SSB states can coexist at the first-order phase transition points. 

\begin{figure}[tbp]
\centering
\includegraphics[scale=0.5]{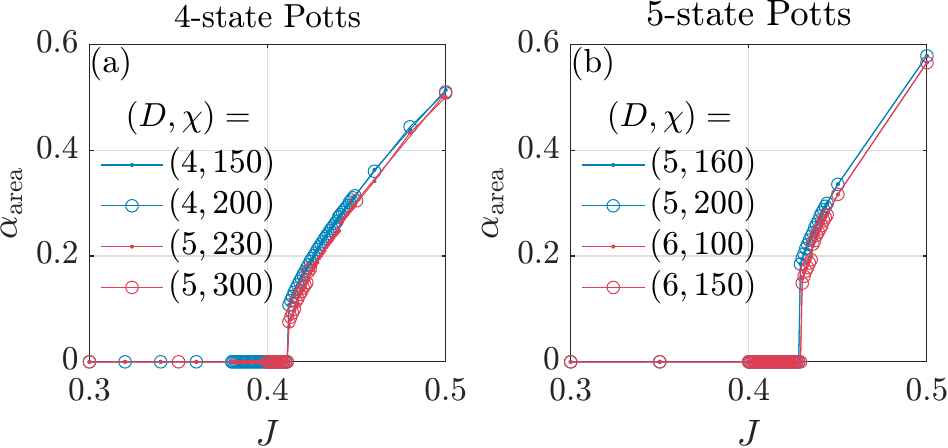}
\caption{Area law coefficients $\alpha_{\tarea}$ of the $(2+1)d$ quantum Potts models from iPEPS with various bond dimension $(D,\chi)$. (a) Results of the 4-state Potts model. (b) Results of the 5-state Potts model.}
\label{Fig:results_Potts_main_text}
\end{figure}

\textbf{Discussions and outlook.}  In this work, we propose an iPEPS framework for disorder parameters. We show that the area and perimeter law coefficients of disorder parameters can be directly extracted from iPEPS without \textcolor{black}{finite system or subsystem size} extrapolation. We propose a \textcolor{black}{finite correlation length} scaling theory for the disorder parameter and validate it using variationally optimized iPEPS. We find that from careful finite scaling analysis, the disorder parameter satisfies the perimeter law when the correlation length $\xi\rightarrow\infty$, which indicates that the higher-form symmetry in the dual lattice gauge model breaks spontaneously at the critical point. We also study a class of first-order phase transitions, at which both the perimeter law and the area law of disorder parameters can coexist. 

The iPEPS method evaluating disorder parameters and the \textcolor{black}{finite correlation length} scaling theory of the disorder parameter can be further extended to systems with continuous global symmetries~\cite{ZY_Memg_disorder_U_1_2021}. Moreover, the disorder parameter can be generalized to various internal symmetries, i.e., higher-form symmetries~\cite{BRICMONT_1983,FM_1983,FM_1986,xu_2024_FM} and subsystem symmetries~\cite{Fracton_2016,zhu_2023}, and such disorder parameters would be useful to detect topological (fracton) phases and their transitions.

\textbf{Acknowledgements.} 
The authors appreciate Bram Vanhecke and Laurens Vanderstraeten for providing the optimized boundary iPEPS tensors of the $3d$ classical Ising model. W.-T. Xu thanks Yu-Jie Liu for helpful discussions and acknowledges support from the Munich Quantum Valley, which is supported by the Bavarian
state government with funds from the Hightech Agenda
Bayern Plus. R.-Z. H thanks Cheng-Xiang Ding, Yun-Long Zang and Frank Verstraete for helpful discussions. R.-Z. H is supported by NSAF No. U2330401, and a postdoctoral fellowship from the Special Research Fund (BOF) of Ghent University.

\textbf{Data availability.} Data, data analysis, and simulation codes are available on Zenodo~\cite{zenodo}.

\bibliography{refs}

\newpage
\clearpage
\appendix
\onecolumngrid

\begin{centering}
{\large \textbf{ Supplemental Materials for \\``\thetitle''}}
\end{centering}
\section{Disorder Parameter and Higher-Form Symmetry}

Before discussing the relation between disorder parameters and higher-form symmetries\textcolor{black}{~\cite{High_form_Kapistin_2015,McGreevy_2023,Bhardwaj:2023}}, let's first briefly review the definition of local order parameters. Consider a system with a global on-site symmetry $U_g=\prod_i U_i(g),g\in G$ where $G$ is a finite group and $U_i(g)$ is a local unitary representation at the site $i$.
The order parameter is the expectation value of a local $G$ symmetry charge operator $O_{\bm{\omega}}$, which transforms according to a one-dimensional non-trivial irrep $\bm{\omega}=\{\omega_g|g\in G\}$ of $G$: $U_g O_{\bm{\omega}} U^{\dagger}_g=\omega_g O_{\bm{\omega}},\forall g$.
In the symmetric phase, it is easy to prove that the order parameter $\langle O_{\bm{\omega}}\rangle$ must be $0$. If $\langle O_{\bm{\omega}}\rangle\neq 0$ in a system with the global symmetry $G$, it implies that the symmetry $G$ breaks spontaneously. 

Using the generalized duality transformation~\cite{Wegner_duality_1971,Levin_Gu_2012}, which is closely related to the concept ``gauging", disorder parameters can be understood as order parameters of higher-form symmetries. Applying the generalized duality transformation to a system with a global on-site symmetry $G$ in $d$-spatial dimensions, one can obtain a $\text{Rep}(G)$ lattice gauge theory, where $\text{Rep}(G)$ is a category of irreps of $G$\textcolor{black}{~\cite{Bhardwaj2018}}. In particular, $\text{Rep}(G)\simeq G$ when $G$ is Abelian, and $\text{Rep}(G)$ is a non-invertible symmetry when $G$ is a non-Abelian group. This lattice gauge theory has a $(d-1)$-form symmetry, which is nothing but a Wilson loop operator $W_{\bm{\omega}}$ with $\bm{\omega}\in \text{Rep}(G)$ in $d$ spatial dimensions, and $W_{\bm{\omega}}$ is an MPO when $\bm{\omega}$ is not a one-dimensional irrep. The charge operator $O_g^{[d-1]}$ of the $(d-1)$-form Wilson loop symmetry applies on the $(d-1)$-dimensional sub-manifold and satisfies $W_{\bm{\omega}} O_g^{[d-1]} W_{\bm{\omega}}^{\dagger}=\omega_g^N O_g^{[d-1]},\forall g$, where $N$ is the number of intersection points between $O_g^{[d-1]}$ and $W_{\bm{\omega}}$. Notice that $O_g$ and $W_{\omega}$ can freely deform as long as their homotopy class does not change. 
The dual of the higher-form symmetry charge operator $O_g^{[d-1]}$ is nothing but the non-local disorder operator $\prod_{i\in A}U_i(g)$, where $A$ is a subsystem in $d$ spatial dimensions and its boundary $\partial A$ is the $(d-1)$ dimensional manifold on which $O_g^{[d-1]}$ lives.

The $G$ symmetric phase of the original model corresponds to the deconfined phase of the Rep($G$) lattice gauge theory, where the higher-form symmetry charge operators satisfy the perimeter law: $\langle O_g^{[d-1]}\rangle\simeq e^{-\alpha_{\tperi} |\partial A|}$ for a sufficient large $A$ with $\alpha_{\tperi}$ being the perimeter law coefficient, and it indicates the $(d-1)$-form Wilson loop symmetry must break spontaneously. So, if the disorder parameter $\tilde{m}_g=\langle\prod_{i\in A}U_i(g)\rangle$ of the original model satisfies the perimeter law, it implies that the phase of the original model must be $G$ symmetric. Meanwhile, the $G$-SSB phase of original model correspond to the Higgs phase of the $G$ lattice gauge theory, where the $(d-1)$-form Wilson loop symmetry does not break and $\langle O_g^{[d-1]}\rangle$ must satisfy the area law  $\langle O_g^{[d-1]}\rangle\simeq e^{-\alpha_{\tarea}  |A|}$\textcolor{black}{~\cite{Wegner_duality_1971,Kogut_1979}}, where $\alpha_{\tarea}$ is the area law coefficient, so the disorder parameter $\tilde{m}_g$ must also satisfy the area law.

\section{Results of Local Physical Observables}
In this section we show conventional local observables, i.e., ground state energy derivatives, local order parameters, and correlation length extracted from the iPEPS for the quantum Ising model and the quantum Potts models as a benchmark of our iPEPS simulation.

Let us first consider the quantum Ising model on the square lattice. The local order parameter $\langle\sigma^z\rangle$ extracted from iPEPS with various bond dimensions is shown in Fig.~\ref{Fig:appendix_quantum_Ising_Potts}(a). Fig.~\ref{Fig:appendix_quantum_Ising_Potts}(b) shows the correlation length extracted from iPEPS with various bond dimensions. In Fig.~\ref{Fig:appendix_quantum_Ising_Potts}(c), we use the known critical exponents $\nu= 0.629970(4)$ and $\beta=0.326418(2)$ of the $(2+1)$D Ising universality class~\cite{Conformal_bootstrap_2016} to perform data collapse of the local order parameter, which matches the prediction from the finite entanglement scaling~\cite{Bram_2022}. Moreover, at the critical point $J_c=0.328474(3)$, we also extrapolate the scaling dimension of the local order parameter, see Fig.~\ref{Fig:appendix_quantum_Ising_Potts}(d), and the obtained result $\Delta_{\sigma}=0.50(2)$ is very close to the (almost) exact results $\Delta_{\sigma}=0.518 1489(10)$~\cite{Conformal_bootstrap_2016}. These results indicate that our iPEPS accurately captures the $(2+1)$D Ising criticality. 

Then we consider the quantum Potts model on the square lattice. The first-order energy density derivative of the 4-state and 5-state ferromagnetic Potts model obtained from iPEPS with various bond dimensions are shown in Figs.~\ref{Fig:appendix_quantum_Ising_Potts}(e) and (f), respectively. The critical point is found to be $J_c = 0.411$ and $J_c=0.429$ for the $q=4$ and $q=5$ models separately. For $q=4$, it is consistent with known results~\cite{quantum_Potts}. Figs.~\ref{Fig:appendix_quantum_Ising_Potts}(g) and (h) show the local order parameters $|\langle Z\rangle|$ of the 4-state and the 5-state Potts models, separately. We can clearly see the discontinuities from the first-order energy density derivatives and the local order parameters, which signal the first-order phase transitions. In addition, the discontinuities get enhanced with the increase of $q$, as expected. 
 
\begin{figure}[b]
\centering
\includegraphics[scale=0.5]{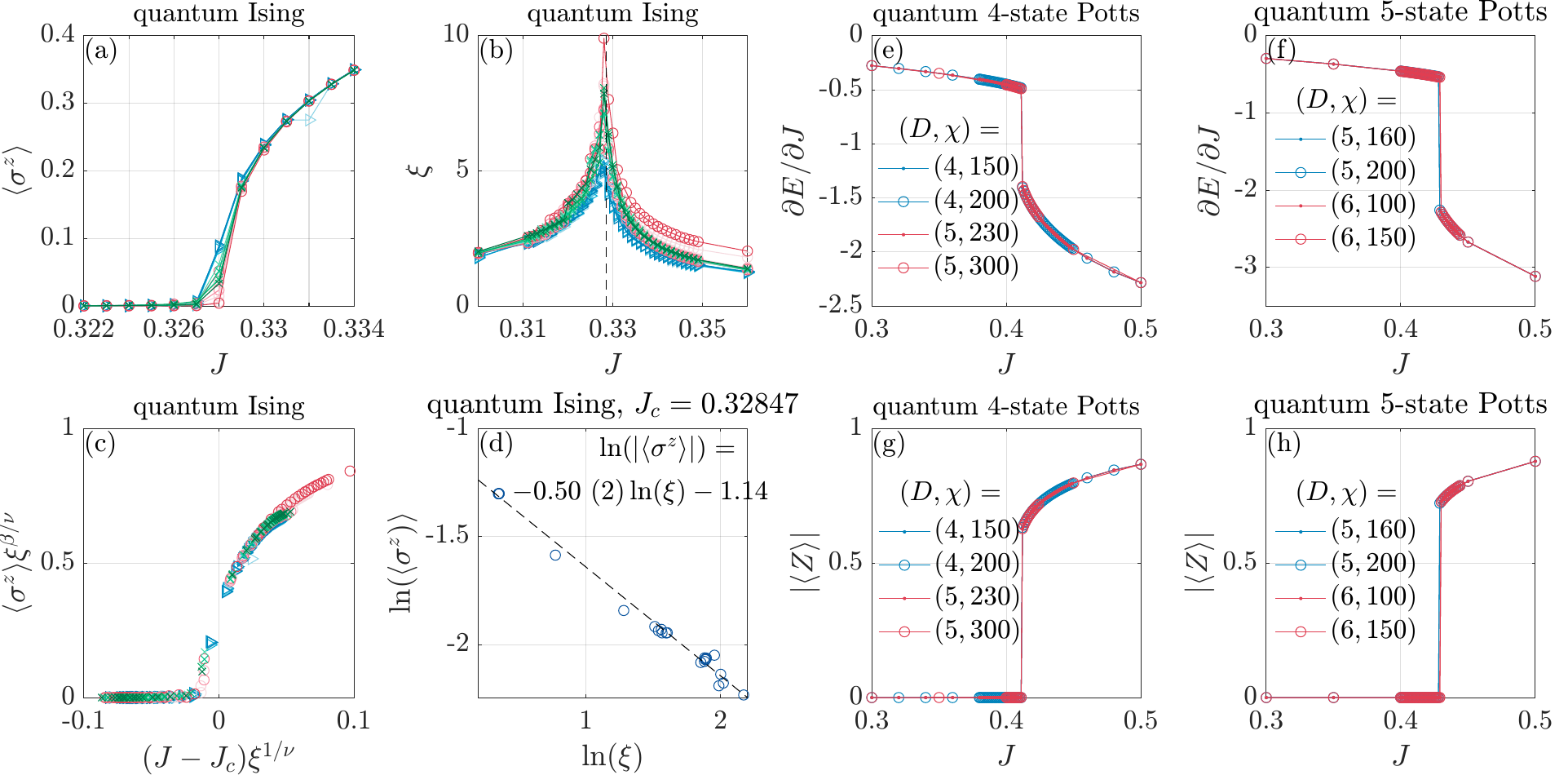}
\caption{Conventional local observables from the iPEPS of the quantum Ising model [(a)-(d)] and the quantum Potts models [(e)-(f)]. Here the legend for the quantum Ising model is the same as that in Fig.~\textcolor{black}{3 in the main text}. (a) The order parameter $\langle \sigma^z \rangle$ around the critical point of the quantum Ising model, obtained from iPEPS with various bond dimensions. (b) The correlation length obtained from iPEPS of the quantum Ising model with various bond dimensions. (c) Data collapse of the order parameter of the quantum Ising model. (d) Double-log plot of the order parameter  $\langle \sigma^z \rangle$ versus the correlation length $\xi$ at the critical point $J_c\approx0.32847$ of the quantum Ising model, and we fit the critical exponent $\beta=0.50(2)$. (e) and (f) show the first-order derivative of the ground state energy of the 4-state Potts model and the 5-state Potts model, respectively. (g) and (h) show the local order parameters of the 4-state Potts model and the 5-state Potts model, respectively.}
\label{Fig:appendix_quantum_Ising_Potts}
\end{figure}

\section{Disorder Parameter of the $3$D Classical Ising Model}
Similar to the simulation of the two-dimensional quantum Ising model, we can perform the same calculation and analysis for the $3$D classical Ising model on the cubic lattice, given by the partition function
\begin{equation}
    \mathcal{Z}(K)=\sum_{\{\sigma^z_i\}} \exp \left(K\sum_{\langle i,j\rangle}\sigma^z_i\sigma^z_j\right),
\end{equation}
where $K$ is the inverse temperature. It is well-known that the model has a continuous phase transition at $K_c=0.2216544(3)$~\cite{Talapov_1996}, described by the $3$D Ising universality class. When $K<K_c$ ($K>K_c$), it is the paramagnetic (ferromagnetic) phase.     

The partition function can be written as a $3$D tensor network, whose transfer matrix of the partition function is a two-dimensional Hermitian infinite projected entangled pair operator (iPEPO). The fixed point of the transfer matrix can be approximated by an iPEPS and optimized variationally~\cite{Laurens_3D_TN_2018}, see code in Ref.~\cite{zenodo}. Given the fixed point iPEPS, we can evaluate the disorder parameters using the method in the main text. 

The results of the perimeter law coefficient $\alpha_{\text{peri}}$ and the area law coefficient $\alpha_{\text{area}}$ of the $3$D classical Ising model are shown in Figs.~\ref{Fig:results_3d_classical_Ising}(a) and (b). We perform data collapse of the perimeter law coefficient $\alpha_{\text{peri}}$ and the area law coefficient $\alpha_{\text{area}}$ in Figs.~\ref{Fig:results_3d_classical_Ising}(c) and (d), using the critical exponent $\nu\approx0.629970(4)$ of the $3$D Ising CFT~\cite{Conformal_bootstrap_2016}. In Fig.~\ref{Fig:results_3d_classical_Ising}(c), we obtain the best data collapse by taking $\alpha_{\text{peri},c}=0.0257$. In Fig.~\ref{Fig:results_3d_classical_Ising}(e), we extrapolate the perimeter law coefficient at the critical point $K_c$ from the finite bond dimension results, and obtain $\alpha_{\text{peri},c}=0.0260(2)$, which is consistent with $\alpha_{\text{peri},c}$ from Fig.~\ref{Fig:results_3d_classical_Ising}(c).  We also   
extrapolate the area law coefficient $\alpha_{\text{area}}$ at the critical point $K_c$ from the finite bond dimension results, see Fig.~\ref{Fig:results_3d_classical_Ising}(b), and obtain $\alpha_{\text{area},c}=0.00003(16)$, implying that at the critical point the disorder parameter satisfies the perimeter law. It also indicates that in the dual $3d$ classical $\mathbb{Z}_2$ lattice gauge model~\cite{Wegner_duality_1971,Kogut_1979}, the 1-form  Wilson loop symmetry breaks spontaneously at the critical point. Notice that the perimeter law coefficient $\alpha_{\text{peri},c}$ at the critical point of the $(2+1)$D quantum Ising model and the one of the $3$D classical Ising model are different, indicating that $\alpha_{\text{peri},c}$ is not a universal quantity. In addition, the results about the local order parameter of the $3$D classical Ising model and its data collapse can be found in Ref.~\cite{TFD_state_2024}.  

\begin{figure*}[tbp]
\centering
\includegraphics[scale=0.5]{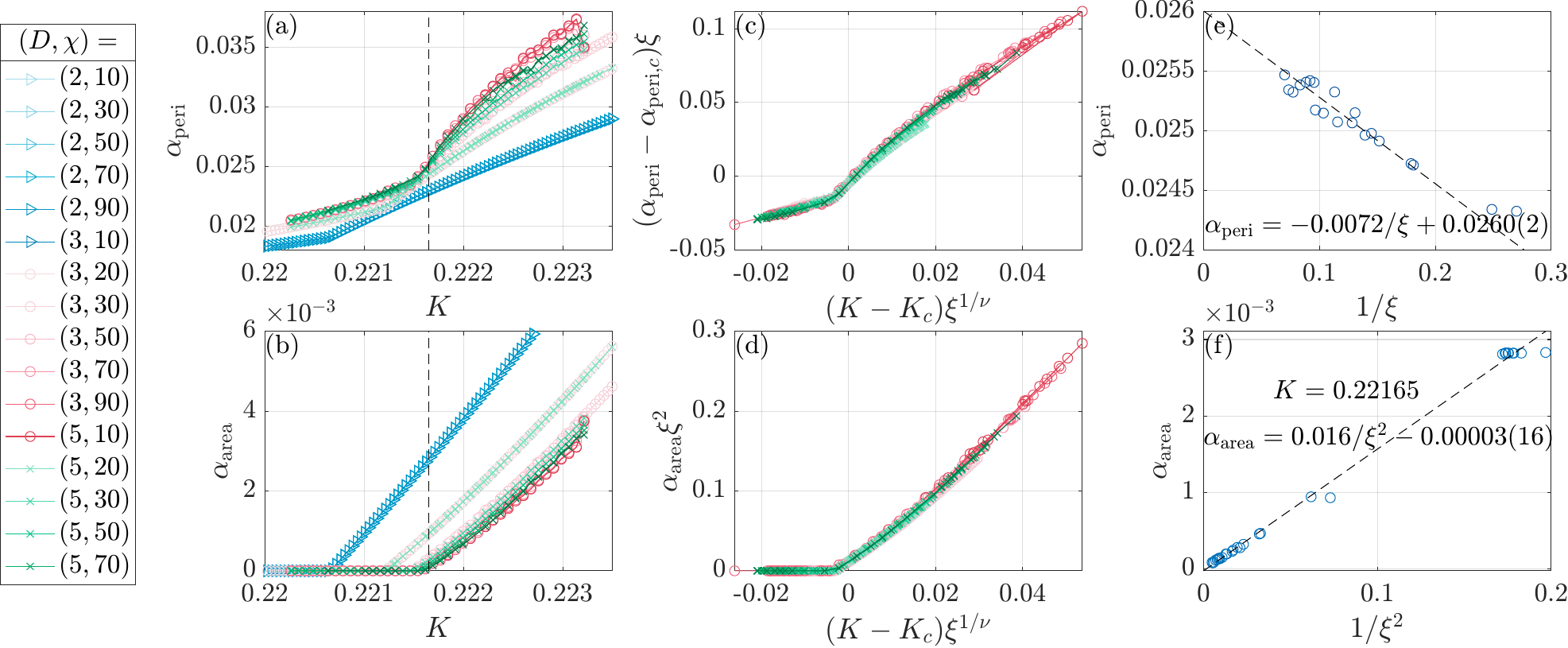}
\caption{Disorder parameter of the $3$D classical Ising model on the cubic lattice. (a) and (b) show the perimeter law coefficient $\alpha_{\tperi}$ and the area law $\alpha_{\tarea}$ coefficient obtained from iPEPS with various bond dimensions $(D,\chi)$, separately. (c) and (d) denote data collapses of the perimeter and area law coefficients after rescaling, respectively. (e) and (f) show the universal scaling behavior of the perimeter and area law coefficients, respectively, at the critical point $K_c=0.2216544(3)$. The data are obtained from iPEPS with bond dimensions $D=2,3,4$ and $\chi$ ranging from $10$ to $90$.}
\label{Fig:results_3d_classical_Ising}
\end{figure*}

\section{Vanishing of Area Law Coefficient at a Quantum Critical Point with $z>1$}
In the main text, we showed that the finite entanglement scaling theory applies to a quantum critical point described by the Ising CFT with the emergent Lorentz invariance characterized by a dynamical critical exponent $z=1$. Actually, quantum critical points without emergent Lorentz invariance are also very common. A large class of $z >1$ quantum critical points can be accurately even exactly described by iPEPS with constant bond dimensions~\cite{PEPS_from_classical_2006}. A natural question is which kind of law disorder parameters satisfy at these $z >1$ quantum critical points. To answer this question, we study the disorder parameter at a quantum critical point with \textcolor{black}{$z>1$}.

We consider the model defined on the square lattice~\cite{Near_and_at_CQCP_2011}:
\ie 
H_0 = -\frac{1}{\eta^4}\sum_{i} \left( \sigma^x_i + \frac{1}{\eta^4} \prod_{\langle i,j\rangle} \eta^{\sigma^z_i \sigma^z_j} \right),
\fe
where $\eta$ is a tuning parameter. The model has a global $\mathbb{Z}_2$ symmetry generated by $\prod_i \sigma^x_i$. Its ground state is in a gapped $\mathbb{Z}_2$ symmetric or a $\mathbb{Z}_2$ SSB gapped phase at large or small $\eta$, respectively. There is a quantum critical point with $z \approx 2.167$ at $\eta_c = \left(1+\sqrt{2}\right)$ between these two phases. 

Interestingly, the ground state of this model can be exactly represented by an iPEPS with a bond dimension $D=2$ for all $\eta$. In the SSB phase, there is a subtlety because the exact iPEPS is a cat state, i.e., a superposition of two $\mathbb{Z}_2$ SSB states, such that the broken symmetry is recovered and the disorder parameter of the exact iPEPS satisfies the perimeter law for all $\eta$ even in the SSB phase~\cite{Gaining_insight_2023}. Therefore, it is not convincing to calculate the disorder parameter using the exact iPEPS since at the critical point the iPEPS could also be a cat state. To avoid the subtlety, we add a small relevant symmetry-breaking field $h$ to the Hamiltonian: 
\ie \label{Eq:dfising}
H = H_0(\eta_c) + h \sum_i \sigma_i^z,
\fe
where $h$ induces a finite length scale $\xi$ at $\eta_c$. Using the variational iPEPS, we calculate the disorder parameter at different $h$ and perform a finite-length scaling analysis. As expected, the length scale obtained by the relevant field shows an accurate power-law behavior, see Fig.~\ref{Fig:df_ising}(a). Using the finite length scale $\xi$, one can further study the area law coefficient $\alpha_\text{area}$. Fig.~\ref{Fig:df_ising}(b) shows the area law coefficient satisfies a perfect scaling behavior with an exponent $\Delta_{\alpha_{\text{area}}} \approx 2.03$, which is close to $2$, as expected. At the zero-field limit ($h=0$, i.e., $\xi \rightarrow +\infty$), $\alpha_\text{area}$ is found approximately zero, indicating that the disorder parameter satisfies the perimeter law at the quantum critical point with $z>1$ and the 1-form symmetry of the dual gauge model breaks spontaneously. 

\begin{figure*}[tbp]
\centering
\includegraphics[scale=0.38]{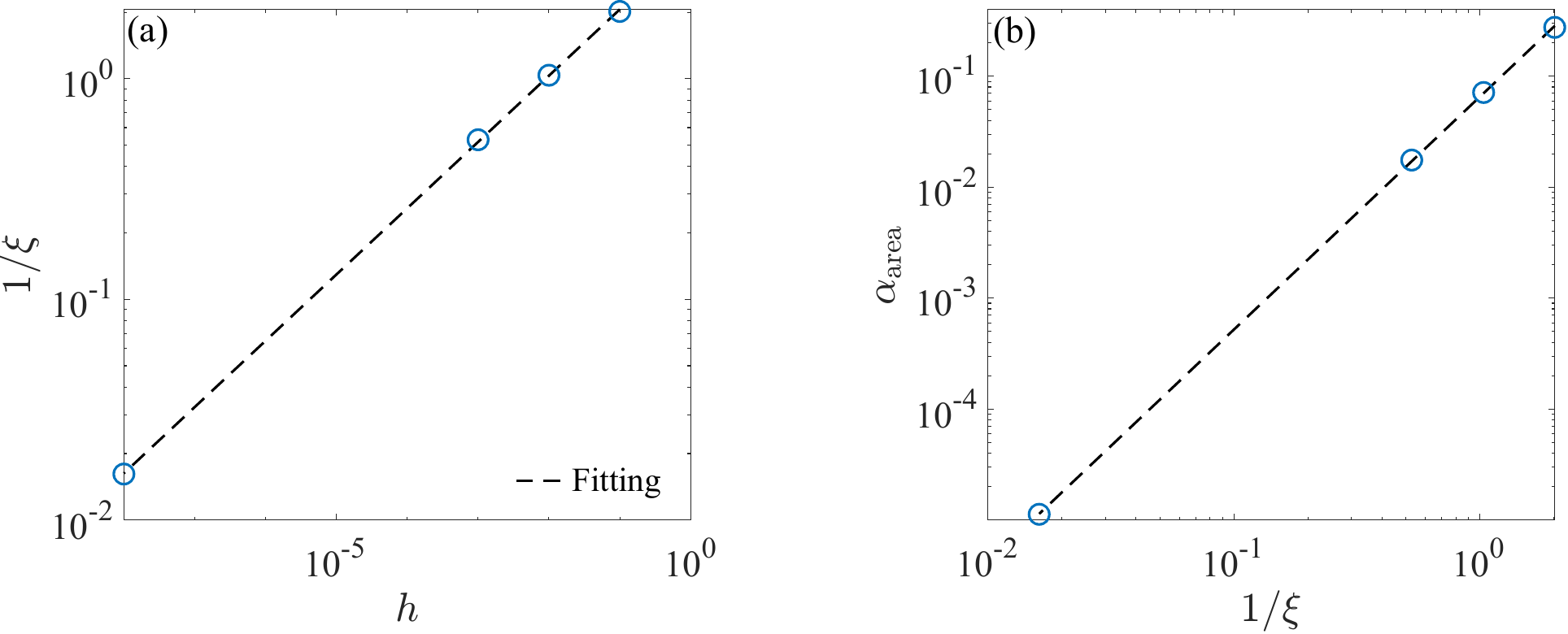}
\caption{Finite-length scaling of the area law coefficient $\alpha_{\tarea}$ in the model Eq.~\eqref{Eq:dfising} obtained from variational iPEPS. The correlation length $\xi$ is obtained by extrapolating bond dimension $\chi$ in the CTMRG to infinity, and the iPEPS bond dimension is $D=2$. (a) Double-log plot of the finite length scale $\xi$ induced by the relevant field $h$. It satisfies $1/\xi \sim h^{1/\Delta_h}$, where the $\Delta_h$ is the scaling dimension of $h$. We have $\Delta_h\approx 3.33$ from the numerical fitting. Notice that the scaling dimension of $\sigma_z$ is $d+z-\Delta_h$.  (b)  Double-log plot of the area law coefficient of the disorder parameter versus the length scale. It satisfies the scaling relation $\alpha_\text{area} \sim a\,\xi^{-\Delta_{\alpha_{\text{area}}}} + \alpha_0$ from Eq.~(7) in the main text with $y=0$ since $\eta=\eta_c$. The exponent is found to be $\Delta_{\alpha_{\text{area}}}= 2.03$ and $\alpha_0$ is found close to zero with a order of magnitude $O({10}^{-6})$. }  
\label{Fig:df_ising}
\end{figure*}

\end{document}